# The discovery of a short period double-degenerate binary star

T. R. Marsh
*University of Southampton, Department of Physics, Highfield, Southampton SO17 1BJ*

**ABSTRACT**
We have found that the white dwarf PG 1101+364 is a double-lined white dwarf/white dwarf binary with an orbital period of 3.47 hours. PG 1101+364 is the shortest period detached double-degenerate yet found and gravitational radiation will cause it to merge in $2.5 \times 10^9$ years. PG 1101+364 has a mass ratio of $0.87 \pm 0.03$. From the relative strengths of the narrow core of H$\alpha$, the lighter and therefore larger star appears to contribute more of the light by an amount that indicates almost equal temperatures for the two stars. We chose to observe PG 1101+364 on the basis of its low spectroscopic mass ($0.31\,M_\odot$), and the discovery of its binary nature confirms the belief that binary evolution is required to produce such low mass white dwarfs. The radial velocity semi-amplitudes ($69.7 \pm 1.7\,\mathrm{km\,s^{-1}}$ and $80.3 \pm 1.6\,\mathrm{km\,s^{-1}}$) and the spectroscopic mass show that we see the orbit of PG 1101+364 at a low inclination of $\approx 25°$.

## 1 INTRODUCTION

It is thought that about 20% of white dwarfs may be born in close binary systems containing two degenerate stars (Iben 1991). However, for some time only two such systems were known (L870−2, $P = 1.56$ days, Saffer, Liebert & Olszewski, 1988, and WD 0957−666, $P = 1.18$ days, Bragaglia et al. 1990). Such systems provide useful constraints upon the common envelope phase of binary evolution during which two stars orbit inside a single envelope. During this stage, which occurs if one star is unable to accrete mass flowing from its companion at a high enough rate, orbital angular momentum and energy is lost to the envelope which is eventually ejected. The loss of angular momentum can have a dramatic effect upon the orbit, shrinking it by two or three orders of magnitude or even forcing a merger of the two stars. The numbers and characteristics of many close binaries depend upon the outcome of the common envelope phase. It is certain that double degenerate binaries have gone through at least one such stage since white dwarfs cannot accrete at high rates.

The common envelope phase is short-lived, and even if a star is caught at the right time, it may not be obvious to the observer. As a result it is better to study the end-products of the common envelope phase of which there are many examples. Many of these have disadvantages. For instance, the cataclysmic variable stars are numerous, but magnetic stellar braking can have a considerable effect upon the orbit, and so it is difficult to know what the orbital parameters were immediately after the common envelope was ejected. The central stars of planetary nebulae avoid this problem, but there are not very many of them and the evolutionary state of the two stars produces difficulties in interpretation. To a large extent double degenerate binary stars avoid these problems although few of them have been found.

Recently we started a program of searching for binary stars amongst low mass ($< 0.45\,M_\odot$) white dwarfs on the basis that such stars cannot have come from single stars as the Galaxy is not old enough. The targets were obtained from the sample of 129 DA white dwarfs observed by Bergeron, Saffer and Liebert (1992). Of the first seven systems we looked at, five were binary stars (Marsh, Dhillon and Duck 1995). We are continuing this program and this paper describes the discovery of a sixth new system, PG 1101+364(WD 1101+364). An important feature of PG 1101+364 is that it is very clearly double-lined, and thus it is a short period counterpart of the double-lined system L870−2 and it will be possible to place strong constraints upon its evolution.

## 2 OBSERVATIONS

We took spectra of PG 1101+364 with the double-beam spectrograph ISIS on the 4.2m William Herschel Telescope on the island of La Palma in the Canary Islands. We observed on the nights of 20/21, 21/22 and 24/25 January 1995, taking 3, 7 and 18 spectra respectively. On the blue arm of ISIS we covered 6420 to 6820 at 0.398Å/pixel. The wavelength range had to be centred on the blue side of H$\alpha$ to avoid columns of poor charge transfer on the TEK CCD we used. On the red arm we covered 7850 to 9360 to search for molecular bands as a sign of a low mass main-sequence or brown dwarf companions. The red arm data is not of any interest for PG 1101+364 and we do not present it here.

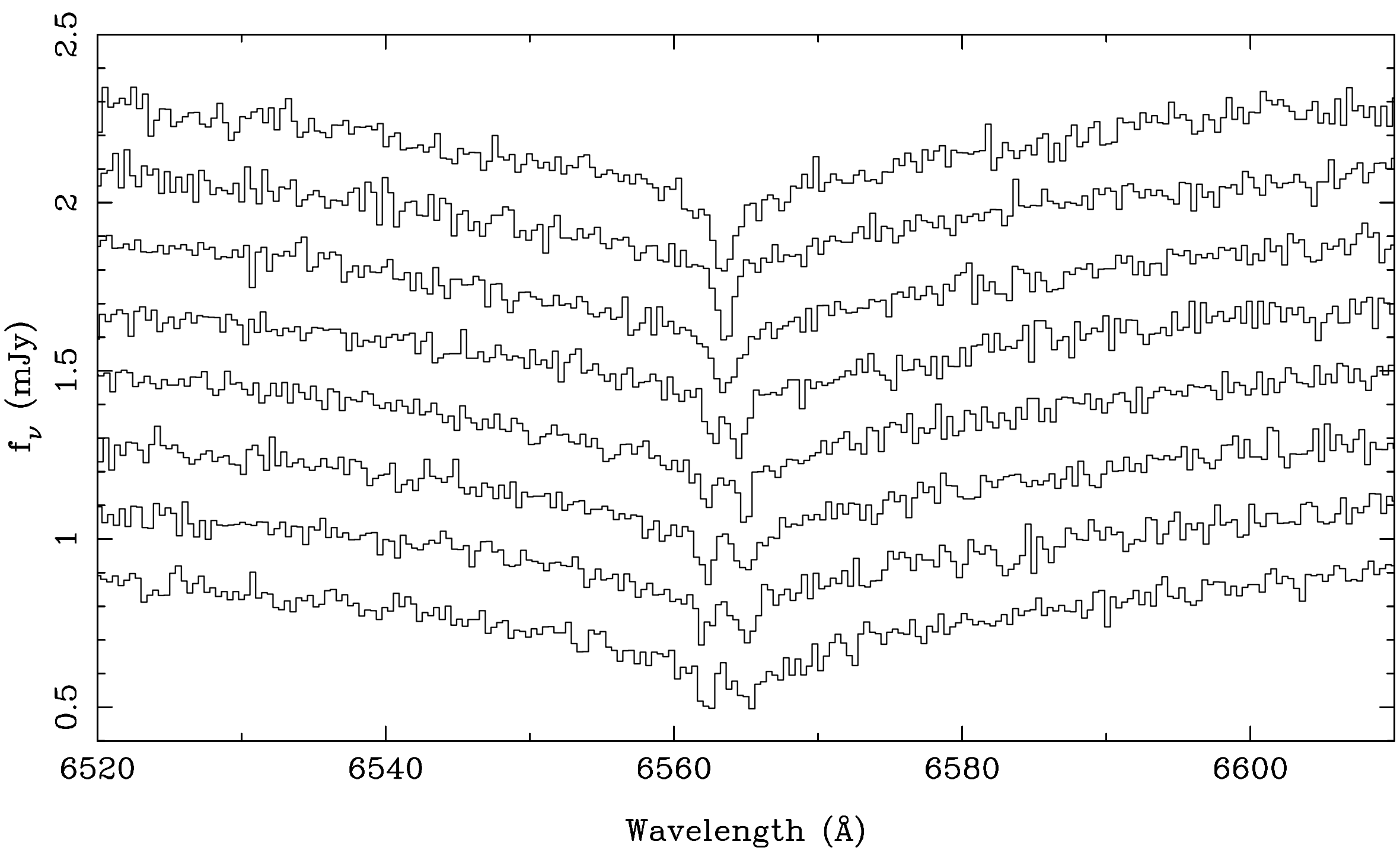

**Figure 1.** The last eight spectra of PG 1101+364 taken as a continuous sequence of 500 second exposures on 25 Jan 1995 show two sharp cores of H$\alpha$ widely separated at the start and merged together by the end. The spectra have been normalised to 1 away from H$\alpha$ and displaced by 0.2 units from each other.

The weather was clear for all the observations of PG 1101+364 although the seeing on the first night was poor (2 to 4 arcsec) compared to the 1 arcsec seeing we enjoyed on the other nights. Most of the exposures were 500 seconds long, with a few longer integrations during periods of poor seeing. The spectra were extracted with weights to give the maximum signal-to-noise. For each object spectrum the arc spectra were extracted at the same position on the detector and then the wavelength scale derived from the arc pair was interpolated in time for the object spectrum. The fits to the arc calibrations had RMS scatters of about 1/30th of a pixel.

## 3 RESULTS

Figure 1 shows the last eight spectra which were taken in a continuous sequence, and it is obvious that PG 1101+364 is a double-lined binary. We measured the radial velocities in the following manner, based upon the methods used by Marsh, Dhillon and Duck (1995) but modified for the double-lined case. We first measured approximate velocities by eye. We then fitted circular orbits to these and obtained fitted velocities for each spectrum. These fitted velocities were then used to offset multiple gaussian fits in which each star was fitted by 4 gaussian components. This fit was made to all 28 spectra simultaneously. The first three of the gaussian components were constrained to have the same full widths at half maximum (FWHM) and peak heights for the two stars; since these components were broad, it was not possible to fit them independently. The fourth component fitted the sharp cores independently for each star. Following this fit, we held the FWHMs and peak heights fixed while we fitted the velocities of each star. The cycle was repeated twice. The data and fitted spectra are displayed in the form of a trailed spectrum in figure 2.

During this process the similarity of the two cores presented an unusual problem because having assigned a given core to one star on a particular night, it was difficult to ensure the same assignment on other nights. Indeed, when measuring by eye at the start we mistakenly reversed the assignment of the first two nights with respect to the third and could only obtain reduced $\chi^2$ of order 3 when fitting the orbits. When we corrected this error the reduced $\chi^2$ dropped to below 1 and there is no doubt about the selection of the correct alias in this case. The heliocentric Julian dates and velocities are listed in table 1 and the phase folded data and circular orbit fits are displayed in figure 3. We have chosen to call the more massive star "star 1" in table 1. In order to measure the relative brightnesses, we made a fit in which we forced the gaussian components of each profile to have the same FWHMs and relative strengths, but allowed the overall strength of each star to vary with respect to each other. On this basis (which is largely determined by the relative strengths of the narrow cores) and assuming that the strength of H$\alpha$ reflects the brightness of the star, star 2 is 1.13 times brighter than star 1.

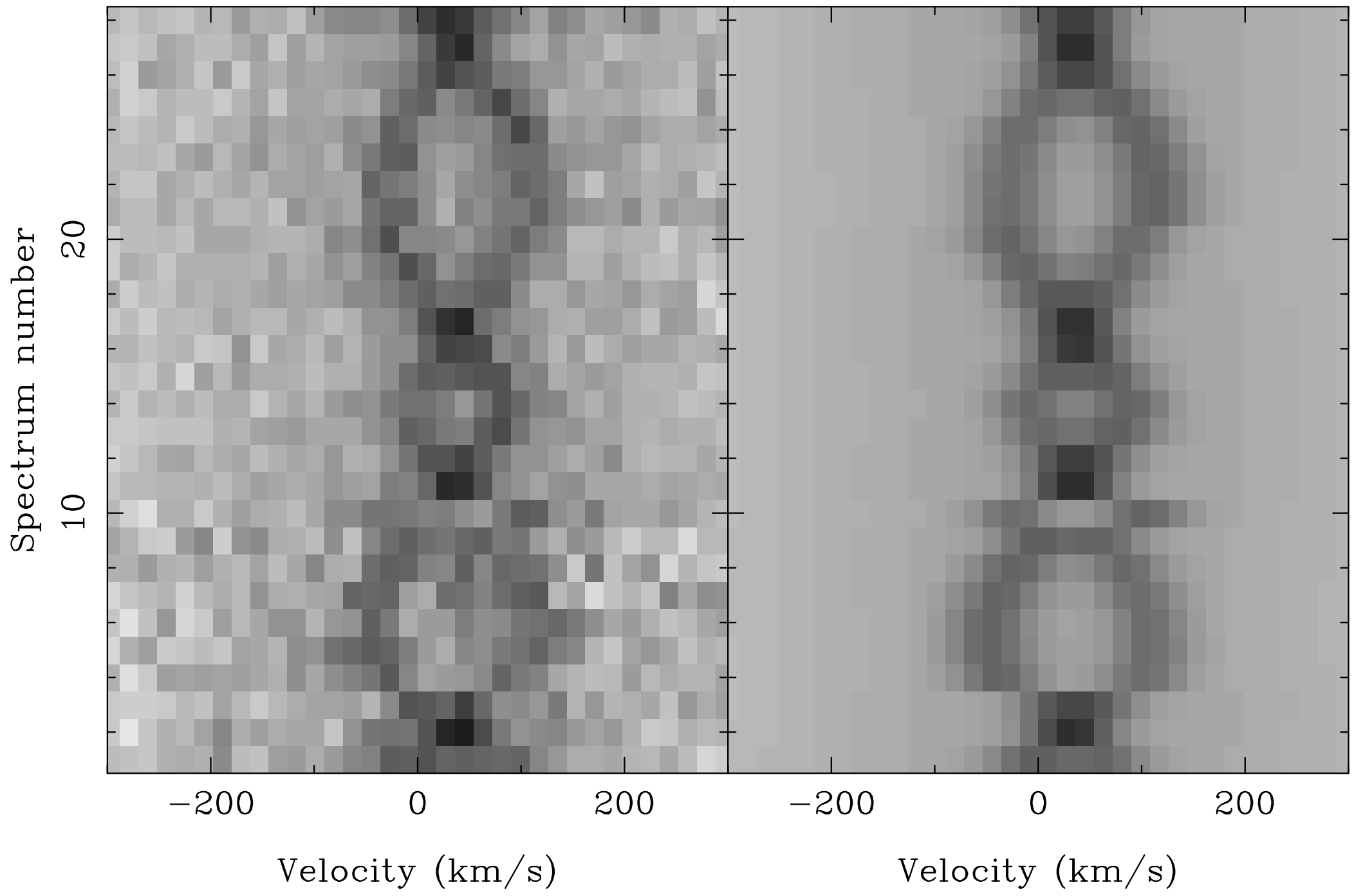

**Figure 2.** In the left panel we plot the data from the 3 nights and in the right panel the spectra calculated from the multi-component fits made in order to derive the velocities. The peculiar pattern here is the result of extensive gaps in phase between groups of spectra.

**Table 1.** Measured velocities

| HJD − 2449700 | Star 1 km s$^{-1}$ | Star 2 km s$^{-1}$ | HJD − 2449700 | Star 1 km s$^{-1}$ | Star 2 km s$^{-1}$ | HJD − 2449700 | Star 1 km s$^{-1}$ | Star 2 km s$^{-1}$ |
|---|---|---|---|---|---|---|---|---|
| 38.66380 | 69.8±6.3 | -8.7±5.9 | 42.57981 | 45.9±6.3 | 30.2±5.9 | 42.75874 | -24.3±5.1 | 107.9±4.6 |
| 38.67230 | 39.9±19.0 | 32.6±17.7 | 42.58693 | 14.3±4.3 | 57.0±4.0 | 42.76467 | -28.3±5.1 | 102.2±4.6 |
| 38.68165 | 5.7±5.3 | 55.9±5.0 | 42.59401 | -0.5±4.3 | 79.5±4.0 | 42.77058 | -24.9±4.9 | 101.9±4.4 |
| 39.63951 | 87.2±6.6 | -38.0±6.1 | 42.63580 | -6.8±4.8 | 91.2±4.5 | 42.77651 | -20.4±4.9 | 96.7±4.4 |
| 39.64663 | 104.9±7.1 | -50.6±6.5 | 42.64171 | 10.7±4.9 | 70.6±4.5 | 42.78397 | -1.8±5.3 | 79.7±4.8 |
| 39.65374 | 124.3±7.7 | -44.3±7.3 | 42.64764 | 20.4±4.4 | 57.2±4.1 | 42.78990 | 9.9±4.8 | 59.3±4.5 |
| 39.66239 | 89.7±9.0 | -42.0±8.6 | 42.65501 | 39.5±10.2 | 29.9±9.6 | 42.79583 | 37.9±25.7 | 33.8±23.8 |
| 39.66951 | 83.5±6.9 | -17.0±6.6 | 42.66094 | 68.7±4.8 | -4.4±4.5 | 42.80175 | 49.4±5.3 | 19.2±4.9 |
| 39.67672 | 67.6±7.1 | -8.6±6.7 | 42.66686 | 80.2±4.7 | -11.1±4.4 | | | |
| 39.71200 | -22.3±5.4 | 107.1±4.9 | 42.67278 | 94.6±4.6 | -29.8±4.3 | | | |

We fitted independent circular orbits to the two stars of the form

$$V = \gamma + K \sin \frac{2\pi}{P}(T - T_0),$$

where $\gamma$ is the systemic velocity, $K$ the radial velocity semi-amplitude, $P$ the orbital period and $T_0$ the time at which the star crosses from blue to red of the systemic velocity if $K > 0$. The fits are listed in table 2 with the semi-amplitude of star 2 negative so that the systemic crossing times can be compared directly.

**Table 2.** Circular orbit fits.

| | Star 1 | Star 2 |
|---|---|---|
| $\chi^2$ | 23.0 | 20.5 |
| $\gamma$ (km s$^{-1}$) | $38.5 \pm 1.3$ | $32.5 \pm 1.2$ |
| $K$ (km s$^{-1}$) | $69.7 \pm 1.7$ | $-80.3 \pm 1.6$ |
| $P$ (days) | $0.144719 \pm 0.000056$ | $0.144656 \pm 0.000044$ |
| $T_0 - 2449742$ | $0.65275 \pm 0.00055$ | $0.65286 \pm 0.00045$ |

## 4 DISCUSSION

### 4.1 The masses of the component stars

Since without the orbital inclination we cannot determine the masses of the two stars, perhaps the most important advantage of a double-lined binary is that we can determine the mass ratio, $q = M_2/M_1 = K_1/K_2 = 0.87 \pm 0.03$. In addition from the orbital period and semi-amplitudes we

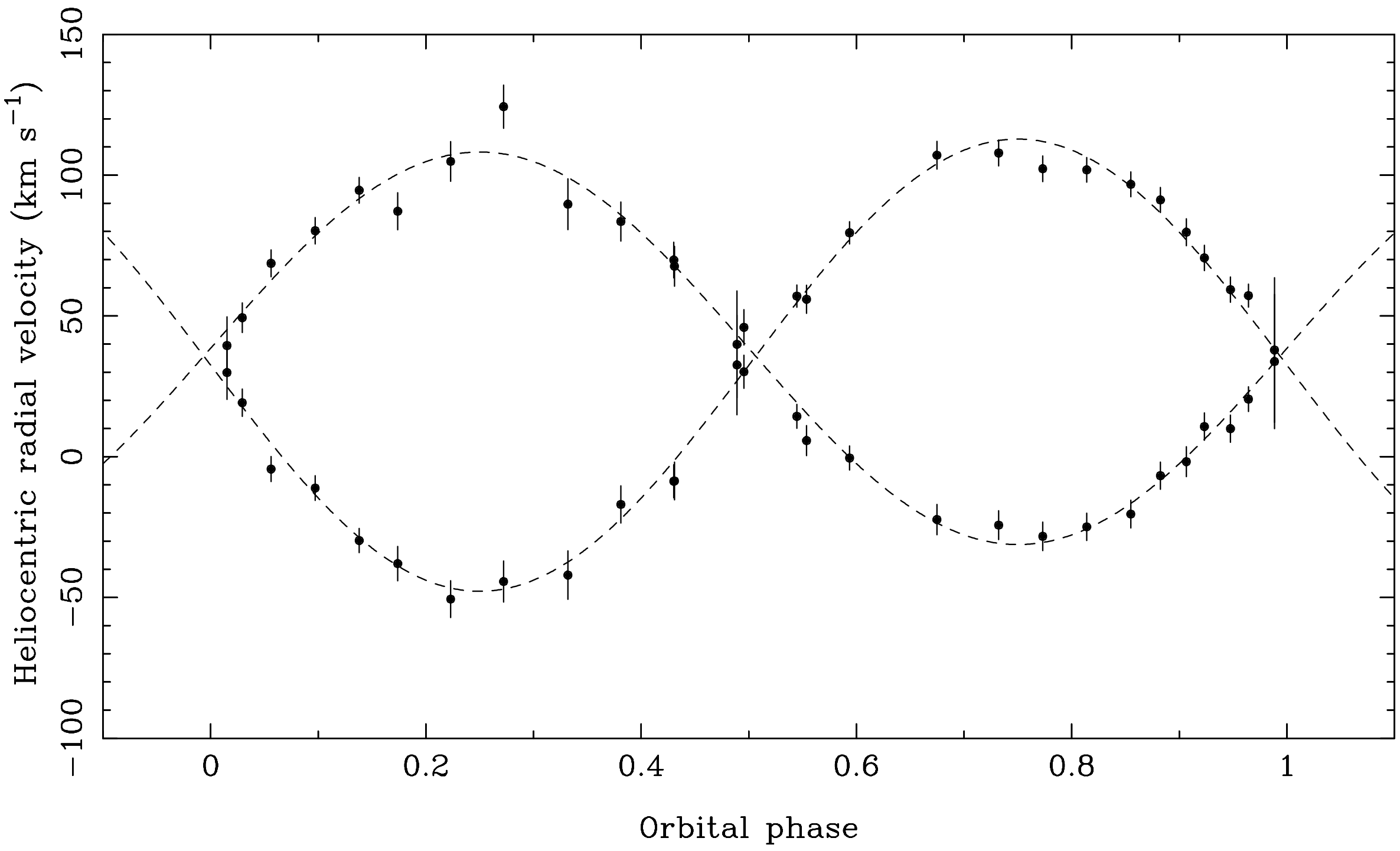

**Figure 3.** The phase folded heliocentric radial velocities and circular orbit fits to PG 1101+364.

can obtain the following quantities:

$$M_1 \sin^3 i = \frac{P}{2\pi G}(K_1 + K_2)^2 K_2,$$
$$M_2 \sin^3 i = \frac{P}{2\pi G}(K_1 + K_2)^2 K_1,$$

where $i$ is the orbital inclination and $M_1$ and $M_2$ are the masses of the two stars. For PG 1101+364 we find

$$M_1 \sin^3 i = 0.0272 \pm 0.0012\, M_\odot$$
$$M_2 \sin^3 i = 0.0236 \pm 0.0012\, M_\odot.$$

Bergeron et al. (1992) measured a mass of $0.31\, M_\odot$, which is presumably a weighted average of the two stellar masses, biassed towards the more luminous and lighter star. If this is correct, then we are seeing the system at an inclination of about 25°.

A further loose constraint comes from the fits of table 2 in which the systemic velocities differ by $\gamma_1 - \gamma_2 = 6.0 \pm 1.8\,\mathrm{km\,s^{-1}}$. The more massive star has the higher systemic velocity which suggests that a difference in gravitational redshift may be significant. For two white dwarfs with masses close to $0.31\, M_\odot$ but differing by a factor of 0.87 we estimate that their gravitational redshifts should differ by $2.2\,\mathrm{km\,s^{-1}}$, reducing the difference to $3.8 \pm 1.8\,\mathrm{km\,s^{-1}}$. This is still marginally significant, and could be taken to favour a mean mass higher than $0.31\, M_\odot$ (e.g. $0.35\, M_\odot$ raises the redshift difference to $2.5\,\mathrm{km\,s^{-1}}$). This is quite possible as Bergeron et al. (1992) derived their masses assuming thin hydrogen layers, whereas results from asteroseismology of ZZ Ceti stars indicate that the hydrogen layers may be thick (Fontaine et al. 1994) which leads to higher masses for a given gravity.

### 4.2 Evolution

If the two stars have the same temperature, then the less massive star, which has the larger radius, should be the brightest. From the mass ratio of 0.87 we estimate that it should be 1.11 times brighter, very close to the factor of 1.13 estimated from H$\alpha$ (this ignores any gravity dependence upon the equivalent width of H$\alpha$). This suggests that the two stars do have a very similar temperature. Since the less massive star should cool faster, it presumably formed after its companion. The mass ratio of $0.87 \pm 0.03$ is close to the mass ratio of L870−2 ($0.90 \pm 0.04$) and agrees with the estimate of $0.87 \pm 0.02$ of Tutukov and Yungelson (1988) for a helium degenerate pair formed from an Algol progenitor.

The future evolution of the binary system is determined by loss of angular momentum through gravitational radiation which occurs at a rate given by (Landau & Lifshitz 1958)

$$\frac{\dot{J}}{J} = -\frac{32}{5}\frac{G^3}{c^5}\frac{M_1 M_2 (M_1 + M_2)}{a^4}.$$

We estimate that the orbital period of PG 1101+364 is currently decreasing on a timescale of $P/\dot{P} = 6.5 \times 10^9$ years and that the two stars will merge within 3/8 of this time, $= 2.5 \times 10^9$ years. This is comparable to the somewhat more massive but longer period system WD 2331+290 found by Marsh, Dhillon and Duck (1995).

On merging the lighter star is accreted onto its companion, and with a total mass less than the Chandrasekhar limit, it is thought that the star is re-born as a helium burning sdO star (Webbink 1984).

## 5 CONCLUSIONS

We have discovered a new short period, detached double-degenerate system PG 1101+364. PG 1101+364 has an orbital period of 3.47 hours and will merge in $2.5 \times 10^9$ years. The spectrum of both components can be seen and we obtain a mass ratio of $0.87 \pm 0.03$. The sharp H$\alpha$ core of the lighter star is stronger than that of its companion by an amount consistent with a similar temperature for the two stars.

## ACKNOWLEDGMENTS

TRM was supported by a PPARC Advanced Fellowship during the course of this work. I thank the referee, P. Bergeron, for helpful comments.